\documentclass[pre,showpacs]{revtex4}
\usepackage[dvips]{graphicx}
\usepackage{graphicx}
\begin{document}
\title{Splitting instability of cellular structures in the Ginzburg-Landau model under the feedback control}
\author{Hidetsugu Sakaguchi}
\affiliation{Department of Applied Science for Electronics and Materials,\\
Interdisciplinary Graduate School of Engineering Sciences,\\
Kyushu University, Kasuga, Fukuoka 816-8580, Japan}
\begin{abstract}
We study numerically a Ginzburg-Landau type equation for micelles in two dimensions. The domain size and the interface length of a cellular structure are controlled by two feedback terms.  The deformation and the successive splitting of the cellular structure are observed when the controlled interface length is increased. The splitting instability is further investigated using  coupled mode equations to understand the bifurcation structure.  
\end{abstract}
\maketitle
 Complicated chemical reaction in a confined cellular structure is considered to be an important step to life.  Oparin considered that  "Coacervate"  played an important role in the origin of a cell in prebiotic chemical evolution~\cite{rf:1}. The compartmentation by some membrane structure is important for a cell to
 be independent of the external atmosphere. We studied the creation and reproduction of model cells with semi-permeable membrane~\cite{rf:2}. Vesicles and micelles can take a cellular form and are considered to be a model system of  primitive  cells~\cite{rf:3,rf:4}. Self-replication of reverse micelles~\cite{rf:5} and vesicles~\cite{rf:6,rf:7} by the increase of the number of the membrane molecules  were observed in experiments. Various types of deformation of the cellular structure such as  budding, splitting and birthing were observed.   
 
On the other hand, the control of spatio-temporal patterns has been an important topic of nonlinear dynamics. The spiral patterns and the spatio-temporal chaos were controlled by some feedback mechanisms~\cite{rf:8,rf:9}. We studied the control of domain size in the Ginzburg-Landau type equation and the method was applied to the problem of cell differentiation~\cite{rf:10,rf:11}. In this paper, we try to control the interface length in a Ginzburg-Landau type model for micelles~\cite{rf:12,rf:13} in two dimensions. We will find a splitting instability of  cellular domains in the model. Although the splitting instability of two-dimensional pulses was observed in the numerical simulation of the Gray-Scott model~\cite{rf:14} and an experiment of the FIS reaction~\cite{rf:15}, it is important  from a viewpoint of artificial cells to study the splitting instability of the cellular structure in such a micelle model. 

Our analysis is based on a free energy functional:
\begin{equation}
F[\phi({\bf r})]=\int d{\bf r}[(c/2)(\nabla^2\phi)^2+(1/2)g(\nabla\phi)^2-(1/2)\phi^2+(1/4)\phi^4-\mu\phi+(1/12)b(\nabla \phi)^4],
\end{equation}
where $\phi$ denotes an order-parameter such as the difference of oil and water concentrations in a problem of the mixture of oil, water and surfactant. The region with a large value of $|\nabla\phi|$ corresponds to the interface region including the surfactant. The surface energy is controlled by parameters $g$ and $b$. 
The larger area (length) of the interface is preferable in case of negative large values of $g$ and $b$.
A time-dependent Ginzburg-Landau equation is given by
\begin{equation}
\frac{\partial \phi}{\partial t}=-\frac{\delta F}{\delta \phi}=\phi-\phi^3+\mu
-c\nabla^4\phi+g\nabla^2\phi+b(\nabla\phi)^2\nabla^2\phi.
\end{equation}
We consider the control of $S_1=\int d{\bf r}\phi$ and $S_2=\int d{\bf r}(\nabla\phi)^2$ in two dimensions. If $c=\mu=b=0$ and $g>0$, there is a domain wall solution 
\begin{equation}
\phi={\rm tanh}(x/\sqrt{2g}).
\end{equation}
Then, $S_1$ is proportional to the domain-size difference of domains satisfying $\phi=1$ and domains satisfying $\phi=-1$, and $S_2$ is approximately expressed as $S_2\sim 4/(3\sqrt{2g})l$, where $l$ is the total length of the interface between the two domains. 
The control of $S_1$ and $S_2$ to certain fixed values leads to the control of the domain size and the interface length. 
If $c$,$\mu$ and $b$ are not zero, the above approximation is not always good, but we call the control of $S_2$  the control of the interface length in this paper. 
We control $S_1$ and $S_2$ to certain fixed values by changing the parameters $\mu$ and $g$ using the global negative feedback as 
\begin{equation}
\frac{d\mu}{dt}=\gamma(S_{10}-S_1),\;\;\frac{dg}{dt}=\gamma(S_2-S_{20}),
\end{equation}
where $\gamma$ is a decay constant for $\mu$ and $g$, and $S_{10}$ and $S_{20}$ are target values of $S_1$ and $S_2$. If $S_1$ is larger (smaller) than the target value $S_{10}$,  $\mu$ decreases (increases), which leads to decrease (increase) $S_1$. 
Similarly, $S_2$ is larger (smaller) than the target value $S_{20}$, $g$ increases (decreases) and the interface region decreases (increases), which leads to decrease (increase) $S_2$. As a result of the negative feedback effect, $S_{1}$ and $S_2$ are expected to approach $S_{10}$ and $S_{20}$. 

We have performed numerical simulation using the pseudo spectral method with Fourier modes of $128\times 128$. The system size is $L\times L=40\times 40$. 
We have assumed the integration range for $S_1$ and $S_2$ as a circular region of radius $L/2=20$ for the comparison with the analysis of coupled mode equations below.  We have obtained qualitatively similar results even if the integration range is assumed to be the total square region of $L\times L$. 
The target value $S_{20}$ is increased slowly from $S_{20}=25$ to $325$. That is, $S_{20}$ is stepwise increased as $S_{20}=25+n$ at $t=100n$ where $n$ is an integer.  The initial condition is assumed to be $\phi=1$ inside of the slightly elliptic region $(x-L/2)^2+1.05(y-L/2)^2=9$, and $\phi=-1$ outside of the region.   
Figures 1(a),(b),(c) and (d) show  cellular domains, where $\phi>0$ is satisfied,  at $S_{20}=50, 55, 65$ and $70$. The domain takes a circular form at $S_{20}=50$. The deformation to an elliptic form begins at $S_{20}\sim 53$, which leads to a dumbbell shape at $S_{20}=65$ and finally it is split into two domains at $S_{20}\sim 68$. The split pattern is observed at $S_{20}=70$ in Fig.~1(d).  
When $S_{20}$ is further increased, more cellular domains appear by the deformation and the splitting. 
Figure 2(a),(b) and (c) shows the cellular domains respectively at $S_{20}=100, 175$ and $275$. The number of cells increases stepwise as 3,4 and 5.
\begin{figure}[t]
\begin{center}
\includegraphics[height=4.5cm]{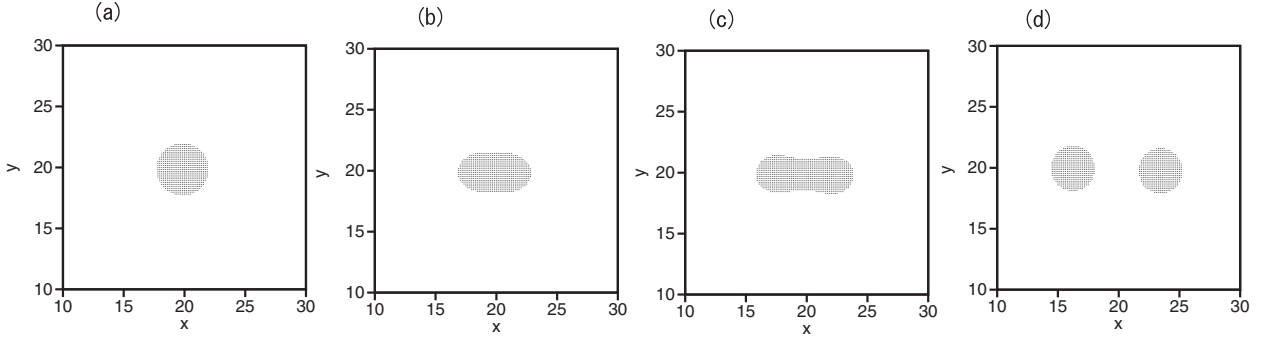}
\end{center}
\caption{Cellular domains satisfying $\phi>0$ at (a) $S_{20}=50$, (b) $55$, (c) 65, and (d) 70.
}
\label{f1}
\end{figure}
\begin{figure}[t]
\begin{center}
\includegraphics[height=4.5cm]{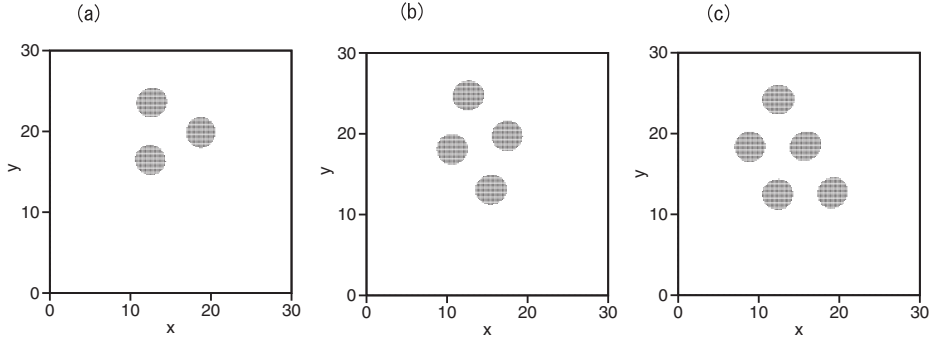}
\end{center}
\caption{Cellular domains satisfying $\phi>0$ at (a) $S_{20}=75$, (b) $100$ and (c) 275.
}
\label{f2}
\end{figure}

The deformation and the splitting instability of a circular domain can be approximately expressed by coupled mode equations.  
From the direct numerical simulation of Eqs.~(2) and (4), it is expected that $\phi({\bf r})$ is approximated at $\phi_0(r,t)+\phi_2(r,t)\cos(2\theta)$, where $r$ and $\theta$ are polar coordinates around the center $(L/2,L/2)$. The substitution of the approximation into Eqs.~(2) and (4) yields 
\begin{eqnarray}
\frac{\partial \phi_0}{\partial t}&=&\phi_0-\phi_0^3-(3/2)\phi_0\phi_2^2+\mu
+b\left \{ \left (\frac{\partial\phi_0}{\partial r}\right )^2+\frac{1}{2}\left (\frac{\partial \phi_2}{\partial r}\right )^2+\frac{2\phi_2^2}{r^2}\right \}\left(\frac{\partial^2 \phi_0}{\partial r^2}+\frac{1}{r}\frac{\partial \phi_0}{\partial r}\right )\nonumber\\
& &+b\frac{\partial \phi_0}{\partial r}\frac{\partial \phi_2}{\partial r}\left (\frac{\partial^2 \phi_2}{\partial r^2}+\frac{1}{r}\frac{\partial \phi_2}{\partial r}-\frac{4\phi_2}{r^2}\right )+g\left(\frac{\partial^2\phi_0}{\partial r^2}+\frac{1}{r}\frac{\partial \phi_0}{\partial r}\right )-\left (\frac{\partial^4\phi_0}{\partial r^4}+\frac{2}{r}\frac{\partial^3\phi_0}{\partial r^3}-\frac{1}{r^2}\frac{\partial^2\phi_0}{\partial r^2}+\frac{1}{r^3}\frac{\partial \phi_0}{\partial r}\right ),\\
\frac{\partial \phi_2}{\partial t}&=&\phi_2-(3/4)\phi_2^3-3\phi_0^2\phi_2+b\left \{ \left (\frac{\partial\phi_0}{\partial r}\right )^2+\frac{1}{2}\left (\frac{\partial \phi_2}{\partial r}\right )^2+\frac{2\phi_2^2}{r^2}+\frac{1}{4}\left (\frac{\partial \phi_2}{\partial r}\right )^2-\frac{\phi_2^2}{r^2}\right \}\left(\frac{\partial^2 \phi_2}{\partial r^2}+\frac{1}{\partial r}\frac{\partial \phi_2}{\partial r}-\frac{4\phi_2}{r^2}\right )\nonumber\\
& &+2b\frac{\partial \phi_0}{\partial r}\frac{\partial \phi_2}{\partial r}\left (\frac{\partial^2 \phi_0}{\partial r^2}+\frac{1}{r}\frac{\partial \phi_0}{\partial r}\right )+g\left(\frac{\partial^2\phi_2}{\partial r^2}+\frac{1}{r}\frac{\partial \phi_2}{\partial r}\frac{4\phi_2}{r^2}\right )-\left (\frac{\partial^4\phi_2}{\partial r^4}+\frac{2}{r}\frac{\partial^3\phi_2}{\partial r^3}-\frac{9}{r^2}\frac{\partial^2\phi_2}{\partial r^2}+\frac{9}{r^3}\frac{\partial \phi_2}{\partial r}\right )
\end{eqnarray}
and 
\begin{eqnarray}
\frac{d\mu}{dt}&=&\gamma \left (S_{10}-\int_0^{L/2}dr 2\pi r\phi_0(r)\right ),\nonumber\\
\frac{dg}{dt}&=&\gamma \left (\int_0^{L/2}dr\{(\partial\phi_0/\partial r)^2+(1/2)(\partial \phi_2/\partial r)^2+2\phi_2^2/r^2\}-S_{20}\right ).
\end{eqnarray}
Note that $\phi_2$ must behave $\phi_2\sim a_2r^2+a_3r^3+\cdots$ near $r=0$, because the angle dependence is $\cos(2\theta)$. 
There exists always a solution with the circular symmetry satisfying $\phi_2=0$. Such a circular solution $\phi_0(r)$ obeys the equations: 
\begin{eqnarray}
\frac{\partial \phi_0}{\partial t}&=&\phi_0-\phi_0^3-(3/2)\phi_0\phi_2^2+\mu
+b\left (\frac{\partial\phi_0}{\partial r}\right )^2\left(\frac{\partial^2 \phi_0}{\partial r^2}+\frac{1}{r}\frac{\partial \phi_0}{\partial r}\right )+g\left(\frac{\partial^2\phi_0}{\partial r^2}+\frac{1}{r}\frac{\partial \phi_0}{\partial r}\right )\nonumber\\
& &-\left (\frac{\partial^4\phi_0}{\partial r^4}+\frac{2}{r}\frac{\partial^3\phi_0}{\partial r^3}-\frac{1}{r^2}\frac{\partial^2\phi_0}{\partial r^2}+\frac{1}{r^3}\frac{\partial \phi_0}{\partial r}\right ),\nonumber\\
\frac{d\mu}{dt}&=&\gamma\left (S_{10}-\int_0^rdr2\pi r\phi_0(r)\right),\nonumber\\
\frac{dg}{dt}&=&\gamma\left (\int_0^rdr(\partial\phi_0/\partial r)^2-S_{20}\right ).
\end{eqnarray}
These equations are obtained from Eqs.~(5) and (7) by setting $\phi_2$ to be zero. 
The linear stability of the circular solution can be investigated by the linear  equation obtained from Eq.~(6):
\begin{eqnarray}
\frac{\partial \phi_2}{\partial t}&=&\phi_2-3\phi_0^2\phi_2+b\left (\frac{\partial\phi_0}{\partial r}\right )^2\left(\frac{\partial^2 \phi_2}{\partial r^2}+\frac{1}{\partial r}\frac{\partial \phi_2}{\partial r}-\frac{4\phi_2}{r^2}\right )+2b\frac{\partial \phi_0}{\partial r}\frac{\partial \phi_2}{\partial r}\left (\frac{\partial^2 \phi_0}{\partial r^2}+\frac{1}{r}\frac{\partial \phi_0}{\partial r}\right )\nonumber\\
& &+g\left( \frac{\partial^2\phi_2}{\partial r^2}+\frac{1}{r}\frac{\partial \phi_2}{\partial r}\frac{4\phi_2}{r^2}\right )-\left (\frac{\partial^4\phi_2}{\partial r^4}+\frac{2}{r}\frac{\partial^3\phi_2}{\partial r^3}-\frac{9}{r^2}\frac{\partial^2\phi_2}{\partial r^2}+\frac{9}{r^3}\frac{\partial \phi_2}{\partial r}\right ).\nonumber\\
\end{eqnarray}
\begin{figure}[t]
\begin{center}
\includegraphics[height=4.5cm]{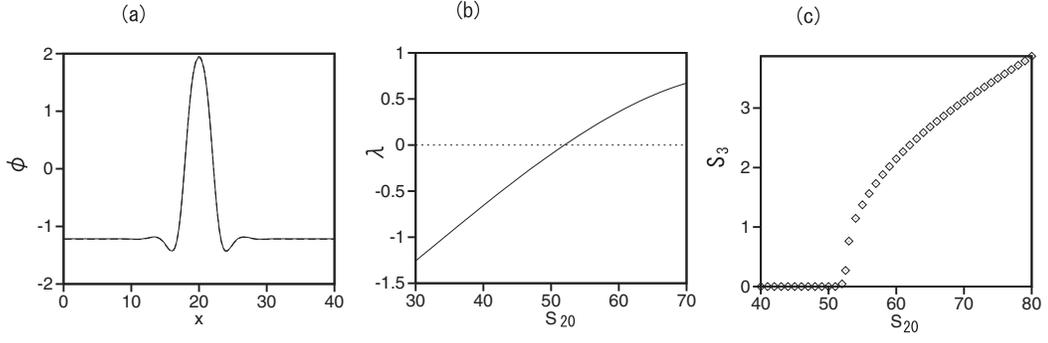}
\end{center}
\caption{(a) Profile of $\phi(x,y)$ (solid curve) at $y=L/2$ for $S_{10}=-1500$ and $S_{20}=50$ in Eqs.~(2) and (4). Profile of $\phi(x)$ by Eq.~(8) ($r$ is set to be $x$.) and the mirror image for $x<0$ (dashed curve) at the same parameters. The two curves are well overlapped. 
(b) Eigenvalue for Eq.~(9) as a function of $S_{20}$ for $S_{10}=-1500$. (c) Mean amplitude $S_3$ of the perturbation $\phi_2(r)$ as a function of $S_{20}$ at $S_{10}=-1500$ for Eqs.~(5),(6) and (7).}
\label{f3}
\end{figure}
\begin{figure}[t]
\begin{center}
\includegraphics[height=4.5cm]{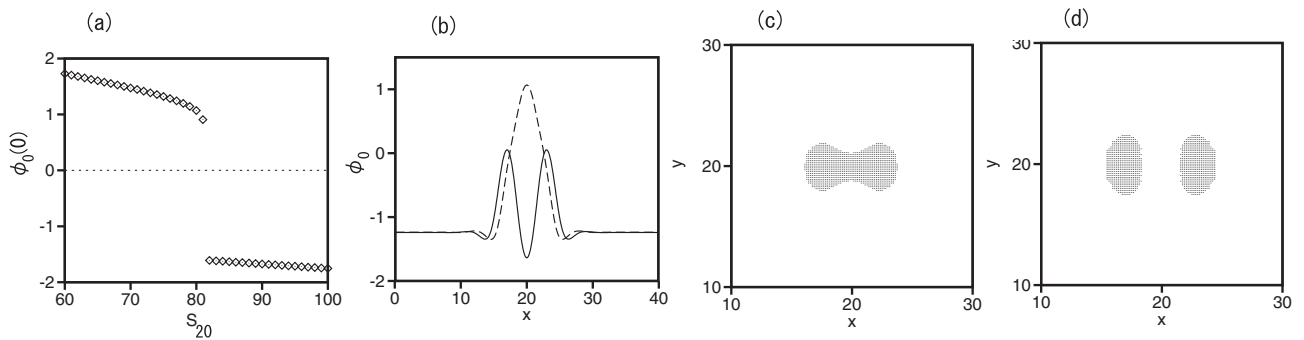}
\end{center}
\caption{(a) $\phi_0(0)$ as a function of $S_{20}$ for $S_{10}=-1500$.  (b)Profile of $\phi_0(x)$ (solid curve) and the mirror image for $x<0$ at $S_{20}=85$ (solid curve) and $S_{20}=80$ (dashed curve).
(c) Deformed domain at $S_{20}=80$. In the shaded domain,  $\phi_0(r)+\phi_2(r)\cos(2\theta)>0$.  (d) Split domains at $S_{20}=85$. In the shaded two domains,  $\phi_0(r)+\phi_2(r)\cos(2\theta)>0$.}
\label{f4}
\end{figure}
 The solid curve in Fig.~3(a) shows the profile of $\phi(x,y)$ at $y=L/2$ at $S_{10}=-1500$ and $S_{20}=50$ for Eqs.~(2) and (4).     
The dashed curve in Fig.~3(a) is a stationary solution to Eq.~(8). 
The two curves are well overlapped, and the difference is hardly seen. That is, the approximation by Eq.~(8) is good.  Figure 3(b) shows the eigenvalue of the linear Eq.~(9)  as a function of $S_{20}$ for a fixed value of $S_{10}=-1500$. The instability occurs at $S_{20}=52$, which is consistent with the critical value $S_{20}\sim 53$  by the direct numerical simulation of Eqs.~(2) and (4).  
Figure 3(c) displays $S_3=\{\int_0^{L/2} drr\phi_2^2\}^{1/2}$ as a function of $S_{20}$ obtained numerically for Eqs.~(5),(6) and (7). It means that the supercritical bifurcation occurs at $S_{20}\sim 52$. That is, the elliptic deformation grows continuously.  The splitting instability is also approximately described by Eqs.~(5),(6) and (7). Figure 4(a) displays the value $\phi_0(0)$ at $r=0$ as a function of $S_{20}$ for $S_{10}=-1500$ obtained by numerical simulation of Eqs.~(5), (6) and (7). A discontinuous transition occurs at $S_{20}\sim 82$. The profile of $\phi_0(r)$ has a peak at $r=0$ for $S_{20}<82$. On the other hand, $\phi_0(r)$ has a peak at nonzero $r$ for $S_{20}>82$ as shown in Fig.~4(b).   
Figure 4(b) displays the profiles of $\phi_0(x)$ at $S_{20}=80$ and 85. The discontinuous transition of the profile $\phi_0(x)$ is clearly seen.  
Figures 4(c) and (d) show the deformation of the cellular domain at (c) $S_{20}=80$ and (d) $S_{20}=85$. In the shaded regions, $\phi_0(r)+\phi_2(r)\cos(2\theta)>0$. The two-peak structure shown in Fig.~4(b) appears as a two-cell structure in Fig.~4(d).   The critical value $S_{20}\sim 82$ in the coupled mode equations is larger than the critical value $S_{20}\sim 68$ of the splitting instability by the direct numerical simulation by Eqs.~(2) and (4). It is partly because the higher modes including $\cos(2m\theta)$ with $m\ge 2$ is truncated in Eqs.~(5),(6) and (7). 
   
In summary, we have proposed a Ginzburg-Landau type model for micelles under the control of the domain size and the interface length. 
As the interface length is increased, a circular cell is deformed to an elliptic form and then split into two cells. By increasing further the interface length, many cells are created by the deformation and the splitting instability. 
We have proposed coupled two-mode equations and found that there are two successive bifurcations for the splitting instability. One is the supercritical bifurcation, where the circular symmetry is broken continuously. At the second bifurcation point, the splitting of the cellular structure occurs discontinuously. In the problem of micelles, we can interpret that the increase of the interface length corresponds to the increase of the surfactant materials created by some chemical reactions inside of the micelles. The splitting processes might be interpreted to correspond to the self-replication process of micelles found in the experiments~\cite{rf:5}.

\end{document}